\documentclass[pra,final,twocolumn,showpacs,showkeys]{revtex4}
\usepackage{amsmath}
\usepackage{graphicx}
\usepackage{amsfonts}
\usepackage{amssymb}
\usepackage{subfigure}
\begin{document}

\title{Relativistic quantum walks}

\author{Frederick W. Strauch}
\email[Electronic address: ]{frederick.strauch@nist.gov}
\affiliation{National Institute of Standards and Technology, Gaithersburg, Maryland 20899-8423, USA}

\date{\today}

\begin{abstract}
By pursuing the deep relation between the one-dimensional Dirac equation and quantum walks, the physical role of quantum interference in the latter is explained.  It is shown that the time evolution of the probability density of a quantum walker, initially localized on a lattice, is directly analogous to relativistic wave-packet spreading.  Analytic wave-packet solutions reveal a striking connection between the discrete and continuous-time quantum walks.
\end{abstract} 
\maketitle

The ``quantum random walk,'' first coined by Aharonov {\it et al.} \cite{Aharonov93}, is a quantum generalization of the classical random walk.  Consider a walker moving on a one-dimensional lattice, taking steps left or right based on the state of a coin.  Classically, if the coin is flipped after each step, this generates a diffusive random walk.  If the coin is quantum mechanical, however, it can be put into a superposition, and rotated by applying a fixed unitary operator.  Aharonov {\it et al.} showed that this quantum procedure (or algorithm) can generate displacements that, on average, are much greater than the classical random walk.

This {\textit{discrete-time quantum walk}} (DTQW) has been re-discovered and extensively analyzed in the context of quantum computation \cite{Kempe2003}.  Two key properties are the following: (i) the standard deviation of the walker's position grows linearly in time ($(\Delta x)_t \sim t$), in clear distinction to the classical random walk ($(\Delta x)_t \sim t^{1/2}$), (ii) for proper initial conditions the walker spreads out symmetrically, with a nearly constant probability distribution save for two peaks located at $x_{\pm} = \pm c t$ (where $c = 1/\sqrt{2}$ for the ``Hadamard walk'' \cite{Kempe2003}), beyond which the probability quickly goes to zero.  

An entirely different approach to ``quantizing'' random walks was initiated by Farhi and Gutmann \cite{Farhi98}.  Beginning with the differential equation for diffusion on a lattice, they performed an analytic continuation to yield a Schr{\"o}dinger equation with a finite-difference Laplacian operator.  This {\textit{continuous-time quantum walk}} (CTQW) was used by Childs {\it et al.} \cite{Childs2003} to construct a special search algorithm that is exponentially faster than classical methods.  Other local search algorithms (with square-root speedup) have been studied using both the discrete and continuous-time quantum walks, often with similar results \cite{Shenvi2003,Childs2004b}.  However, to this author's knowledge, no physical explanation has been proposed to explain the similar performance of these two quantum walks.  

Before connecting these two walks, recall the connection between the DTQW and the Dirac equation.  As discussed by Meyer \cite{Meyer96}, this goes back to Feynman's ``checkerboard,'' a discrete space-time path integral that, in the continuum limit, generates the propagator for the Dirac equation in one dimension \cite{FeynmanBook}.  This is best seen in the following unitary representation \cite{BBirula94}, in which the DTQW is written as the discrete mapping:
\begin{equation}
\left(\begin{array}{l} \psi_R(n,\tau+1) \\ \psi_L(n,\tau+1) \end{array}\right) = U \left(\begin{array}{l} \psi_R(n,\tau) \\ \psi_L(n,\tau) \end{array} \right),
\label{xwalk1}
\end{equation}
$\psi_R$ and $\psi_L$ are wave functions on an infinite lattice, and $U$ is the product of a conditional translation operator and a spin rotation
\begin{equation}
U = [\frac{1}{2} (I+\sigma_z) D + \frac{1}{2}(I-\sigma_z) D^{-1}] e^{- i \theta \sigma_x}.
\label{xwalk2}
\end{equation}
Here the Pauli matrices $\{I,\sigma_x,\sigma_z\}$ act on the spinor components, and the translation operator $D$ acts on wave functions as $(D \psi)(n) = \psi(n-1)$.  The continuum limit is found by letting the position $x = n \epsilon$, $D = e^{-i \epsilon p}$ ($p$ is the momentum), $\theta = m \epsilon$ ($m$ is the mass), and the time $t = \epsilon \tau$. Using the Trotter formula, the limit $\epsilon \to 0$ (with $p, m$, and $t$ finite) yields
\begin{equation}
U^\tau = [ e^{-i \epsilon \sigma_z P} e^{-i \epsilon m \sigma_x}]^{t/\epsilon} \to e^{-i H_{D} t},
\end{equation}
where $H_D$ is the Hamiltonian for the one-dimensional Dirac equation (with $\hbar = c = 1$, $p = -i \partial_x$) \cite{ThallerBook}:
\begin{equation}
i \partial_t \Psi(x,t) = H_{D} \Psi(x,t) = (-i \sigma_z \partial_x + \sigma_x m) \Psi(x,t).
\label{dirac}
\end{equation}

While quite elegant, the properties of this continuum limit have been largely ignored in the extensive analysis of the Hadamard walk \cite{Nayak2000} (in which $e^{-i \pi \sigma_y/4} \sigma_z$ is used in place of $e^{-i \theta \sigma_x}$ in (\ref{xwalk2})).  The closest related work is the continuum limit of the Hadamard walk recently found by Knight {\it et al.} \cite{Knight2003}, but this and the corresponding Airy function solutions are significantly different from the Dirac equation.  Another notable work is that of Meyer \cite{Meyer97}, who studied some of the wavelike properties of quantum cellular automata, but not the uniquely relativistic properties explored here.  Understanding these properties may have importance for quantum algorithms; it has already been shown that massless Dirac operators can improve a continuous-time search algorithm \cite{Childs2004b}.  

Here I use explicit solutions of (\ref{dirac}) to illustrate that the quantum-walk probability distribution is analogous to the spreading of a relativistic particle.  The term ``relativistic'' is taken to mean any evolution of a particle with a maximum speed limit.  The same characteristic spreading---both relativistic and nonrelativistic---is found from a new solution to the quantum walk equations (\ref{xwalk1})-(\ref{xwalk2}) {\it without} going to a continuum limit.  Finally, this solution is found to be analytically related to the continuous-time quantum walk, providing a new link between these two relativistic quantum walks. 

First, it is important to note that, using the Heisenberg equations of motion, wave-packet spreading for any dispersion relation $\omega(p)$ can be written as
\begin{equation}
(\Delta x)_t^2 = (\Delta x)_0^2 + (\Delta v)_0^2 t^2,
\end{equation}
where $(\Delta v)_0$ is the standard deviation of the group velocity $v(p) = d \omega(p)/dp$ \cite{Jordan86}.  For the Dirac equation (\ref{dirac}), the dispersion relation is $\omega(p) = \sqrt{p^2 + m^2}$, and thus $v(p) = p (p^2 + m^2)^{-1/2} < c = 1$, i.e. there is a maximum group velocity, which is of course the speed of light.  

While this linear quantum spreading $(\Delta x)_t \sim t$ is universal, the presence of peaks of the probability distribution (at $x_{\pm} = \pm ct$) depends on the initial localization.  To show this, I construct an explicit time-dependent solution of (\ref{dirac}) by the following Fourier representation:
\begin{equation}
\Psi(x,t) = \frac{\mathcal{N}}{2\pi} \int_{-\infty}^{\infty} dp P_+(p) \frac{1}{\sqrt{2}} \left(\begin{array}{l} 1 \\ 1 \end{array} \right) e^{i p x-(a+i t) \omega(p)}.
\label{diracw1}
\end{equation}
The prefactor $P_+(p) \equiv I + H_{D}/\omega(p)$ projects the spinor onto the positive-energy eigenstates of $H_{D}$, while the parameter $a$ in the exponential allows arbitrary localization in position.  The integrals can be done analytically to yield
\begin{equation}
\Psi(x,t) = \frac{m \mathcal{N}}{\pi \sqrt{2}} \left(\begin{array}{l} s^{-1} K_1(m s) [a + i (t+x)] +K_0(m s) \\ s^{-1} K_1(m s) [a + i ( t-x)] + K_0(m s) \end{array} \right),
\label{diracw2}
\end{equation} 
where $s = [x^2 + (a+i t)^2]^{1/2}$, the normalization factor is
\begin{equation}
\mathcal{N} = \sqrt{\pi/2m} \left[K_1(2 m a) + K_0(2 m a)\right]^{-1/2},
\label{diracw3}
\end{equation}
and $K_n$ is the modified Bessel function of order $n$ \cite{AStegun}.  

\begin{figure}
\centering
\includegraphics[width = 8.5 cm]{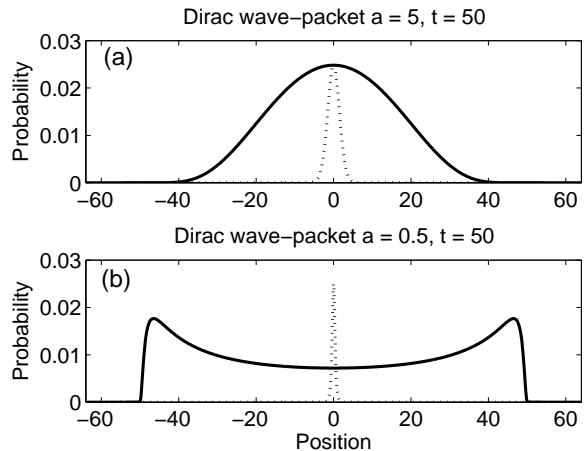}
\caption{(a) Non-relativistic ($a=5$) and (b) relativistic ($a=0.5$) solutions of the one-dimensional Dirac equation.  The probability density $\rho(x,t) = \Psi^{\dagger}(x,t) \Psi(x,t)$ is shown at time $t=0$ (dotted lines, arbitrary units) and at $t=50$ (solid lines).  Other parameters are the mass $m=1$ and the speed of light $c=1$.}
\label{diracfig}
\end{figure}

The probability density is shown in Fig.~\ref{diracfig} for two values of $a$ at $t=0$ and $t = 50$.  The nonrelativistic wave packet (with large $a$) spreads as a Gaussian, while the relativistic wave packet (with small $a$) spreads near the light-cone ($x_{\pm} = \pm c t$) at the speed of light.

Another exact solution of the one-dimensional Dirac equation was found many years ago \cite{Bakke73}.  These examples demonstrate the existence of positive-energy states of a relativistic particle localized beneath its Compton wavelength \cite{Bracken99}, despite well-known claims to the contrary.  Such states appear to require entanglement between the spatial and spinor degrees of freedom \cite{Peres2002}---this is shown for (\ref{diracw2}) below.  

While the resemblance between the probability distribution of the Hadamard walk (see, e.g. \cite{Kempe2003}) and the relativistic wave packet in Fig. 1 is quite strong, it begs the question: what about the non-relativistic case?  For the quantum walk, this requires an initial superposition over the lattice.  A larger spread in the position leads to slower spreading, as expected by the uncertainty principle.  

An analytic solution to the walk equations (\ref{xwalk1}) and (\ref{xwalk2}), covering both the relativistic and nonrelativistic limits, can be found using a similar procedure as above.  I use the Fourier analysis of (\ref{xwalk1}) and (\ref{xwalk2}) and let $D = e^{-i k}$, in which case 
\begin{equation}
U = \left(\begin{array}{ll} e^{-ik} \cos \theta & -i e^{-ik} \sin \theta  \\
-i e^{i k} \sin \theta & e^{i k} \cos \theta \end{array}\right).
\label{coinq1}
\end{equation}
This matrix has eigenvalues $e^{\pm i \omega(k)}$, where $\omega(k)$ satisfies the dispersion relation \cite{Meyer97}
\begin{equation}
\cos \omega(k) = \cos \theta \cos k.
\label{coinq2}
\end{equation} 
The wave packet corresponding to (\ref{diracw1})-(\ref{diracw2}) is
\begin{equation}
\psi(n,\tau) = \frac{N}{2\pi} \int_{-\pi}^{\pi} dk P_+(k) \frac{1}{\sqrt{2}} \left(\begin{array}{l} 1 \\ 1 \end{array}\right) e^{i k n - (\alpha + i \tau) \omega(k)}.
\label{coinqw1}
\end{equation}
The prefactor $P_+(k) \equiv (e^{i \omega(k)} - U)$ projects the spinor onto the ``positive-energy'' eigenstates of $U$, while the parameter $\alpha$ in the exponential allows arbitrary localization on the lattice.  This solution can be written as
\begin{equation}
\psi(n,\tau) = \frac{N}{\sqrt{2}} \left(\begin{array}{l} I_{n}(\tau-1-i \alpha) - e^{-i \theta} I_{n-1}(\tau-i \alpha) \\ I_{n}(\tau-1-i \alpha) - e^{-i \theta} I_{n+1}(\tau-i \alpha) \end{array}\right)
\label{coinqw2}
\end{equation}
where the function $I_n(z)$ is defined by
\begin{equation}
I_n(z) = \frac{1}{2\pi} \int_{-\pi}^{\pi} dk \exp(i k n - i \omega(k) z),
\label{coinqw3}
\end{equation}
with $\omega(k)$ given by (\ref{coinq2}), and the normalization factor is
\begin{equation}
N = [2 I_0(-i 2 \alpha) - e^{i\theta} I_1(-1-i 2 \alpha) - e^{-i\theta} I_1(1-i 2 \alpha)]^{-1/2}.
\label{coinqw4}
\end{equation}

\begin{figure}
\centering
\includegraphics[width = 8.5 cm]{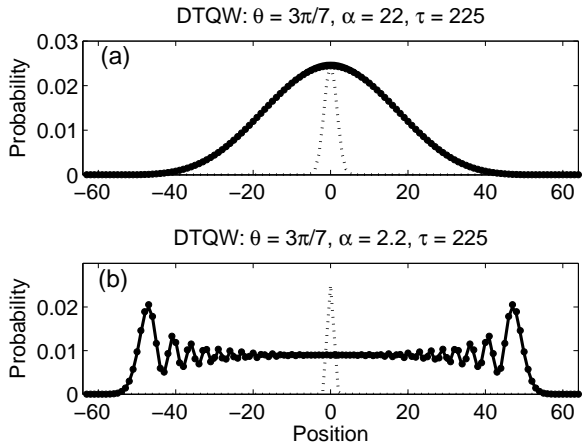}
\caption{(a) Non-relativistic ($\alpha=22$) and (b) relativistic ($\alpha=2.2$) solutions of the one-dimensional discrete-time quantum walk (DTQW). The probability density $\rho(n,\tau) = \psi^{\dagger}(n,\tau) \psi(n,\tau)$ is shown at time $\tau=0$ (dotted lines, arbitrary units) and at $\tau=225$ (dots), along with the Bessel-function approximation (\ref{coinqw5}) (solid lines).  Other parameters are the rotation angle $\theta = 3 \pi/7$, the mass $m=\tan \theta \simeq 4.38$, and the speed of light $c=\cos \theta \simeq 0.22$.  To compare with the Dirac solution, the parameters were chosen such that $c \tau \simeq 50$ and $\alpha/m = \alpha/\tan\theta \simeq a$.}
\label{dtqwfig}
\end{figure}

At this point, a crucial approximation can be made: if $\cos \theta$ is small, replace $\omega(k)$ by its lowest order expansion from (\ref{coinq2}): $\omega(k) \simeq \pi/2 - \cos \theta \cos k$.  This replacement conveniently yields the same maximum group velocity ($\cos \theta$) and allows the following approximation to (\ref{coinqw3}):
\begin{equation}
I_n(z) \simeq e^{i \pi (n-z)/2} J_n(z \cos \theta),
\label{coinqw5}
\end{equation} 
where $J_n$ is the Bessel function of order $n$ \cite{AStegun}.  Using this approximation in (\ref{coinqw2}), as shown in Fig.~\ref{dtqwfig}, compares quite favorably to a numerical calculation of (\ref{xwalk1}) and (\ref{xwalk2}).  The analogy between this and Fig.~\ref{diracfig} is remarkable, taking the ``speed of light'' for the quantum walk as $\cos \theta$.

A few comments are in order.  First, I have shown that evolution on the line for the DTQW, in the relativistic case, has fronts that propagate at the maximum speed $c = \cos \theta$, in close analogy to a solution of the Dirac equation.  Heuristically, the criterion for a relativistic walker is for the the initial localization $(\Delta x)_0$ to be less than the effective Compton wavelength $\lambda = 1/(m c) = 1/\sin\theta$, where the effective mass is given by $m = [d^2 \omega(k) /d k^2]^{-1}_{k=0} = \tan \theta$.  To approximate the Hadamard walk, the appropriate choice is $\theta = \pi/4$, leading to $c = 1/\sqrt{2}$, $m = 1$, and $\lambda = \sqrt{2} > 1$.  Thus, the initial condition most widely studied \cite{Nayak2000}, with the walker localized at one position, has $(\Delta x)_0 \sim 1 < \lambda$, leading to a relativistic quantum walk.  

\begin{figure}
\centering
\includegraphics[width = 8.5 cm]{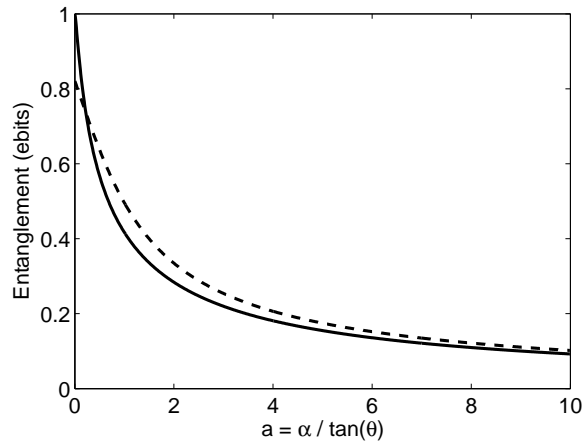} 
\caption{The entanglement, in ebits, of the Dirac solution (solid), and the discrete-time quantum walk with $\theta = 3\pi/7$ (dashed), as a function of the scaling parameter $a = \alpha/\tan \theta$.  The entanglement measure is the spinor entropy \cite{Peres2002}, using a base-2 logarithm.}
\label{entropyfig}
\end{figure}

Second, the analogy between the wave packets of (\ref{diracw2}) and (\ref{coinqw2}) extends beyond the probability distribution to the entanglement between the spinor and spatial degrees of freedom \cite{Peres2002}.  The entanglement as a function of the initial localization is shown in Fig.~\ref{entropyfig}.  By including only positive-frequency terms in the wave function, the entanglement remains constant in time.  Note that here, as in Fig.~\ref{dtqwfig}, I have used the correspondence between the localization parameters $a = \alpha/\tan \theta$, found by comparing the dispersion relations near $k=p=0$. As discussed above, highly localized positive-energy states become significantly entangled in the limit $a \to 0$.

Finally, I note that the particular choice of the spin-rotation (coin) of the quantum walk analyzed above has simplified the calculation.  As an example, the Hadamard walk's dispersion relation $\sin \omega(k) = \sin k / \sqrt{2}$ \cite{Nayak2000} does not satisfy $\omega(k) \simeq \omega(0) + k^2/(2 m)$ for small $k$, but rather $\omega(k) \simeq k /\sqrt{2} - k^3/(12 \sqrt{2})$.  This expansion, the essential approximation used in \cite{Knight2003}, does not lead to an obvious nonrelativistic limit to the Hadamard walk.   

There is, however, both types of propagation---relativistic and nonrelativistic---for the CTQW, defined by \cite{Farhi98}
\begin{equation}
i \partial_t \psi(n,t) = -\gamma \left( \psi(n-1,t) - 2 \psi(n,t) + \psi(n+1,t) \right).
\end{equation}
An exact solution for this walk can be found as above
\begin{equation}
\begin{array}{ll}
\psi(n,t) & = N(2\pi)^{-1} \int_{-\pi}^{\pi} dk e^{i k n - (\alpha+i t) \omega(k)} \\
& = N e^{-2 \gamma(\alpha+i t)} i^n J_n(2 \gamma (t-i \alpha)),
\end{array}
\label{ctqw}
\end{equation}
with the dispersion relation $\omega(k) = 2 \gamma (1-\cos k)$ and normalization factor $N = e^{2 \gamma \alpha} [J_0(-4 i \gamma \alpha)]^{-1/2}$.  The relativistic and nonrelativistic evolution for this case is shown in Fig. 4.  This solution is strikingly similar to (\ref{coinqw2}) [using the relation (\ref{coinqw5})], both visually and analytically, assuming equal maximum speeds $c = 2\gamma = \cos \theta$.  

\begin{figure}
\centering
\includegraphics[width = 8.5 cm]{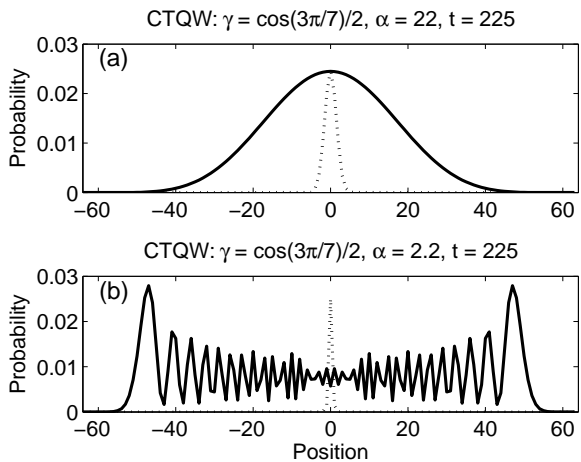}
\caption{(a) Non-relativistic ($\alpha=22$) and (b) relativistic ($\alpha=2.2$) solutions of the one-dimensional continuous-time quantum walk (CTQW). The probability density $\rho(n,t) = \psi^{\dagger}(n,t) \psi(n,t)$ is shown at time $t=0$ (dotted lines, arbitrary units) and at $t=225$ (solid lines). The remaining parameter is $c = 2\gamma = \cos(3\pi/7) \simeq 0.22$. }
\label{ctqwfig}
\end{figure}

A Bessel-function approximation to the DTQW similar to (\ref{coinqw1})-(\ref{coinqw5}) was recently found by an entirely different method \cite{Romanelli2004L}.  The physical content of this approximation, however, is revealed by the wave-packet analysis presented here: when $\cos \theta \ll 1$ ($\theta \sim \pi/2$), the dispersion relations of both the CTQW and DTQW have the common form $\omega(k) = \omega(0) + c (1-\cos k)$, with the ``relativistic'' property of a maximum speed ($v(k) = d \omega/dk < c$).  This equivalence is quite unexpected, since it is the $\theta \ll 1$ ($\cos \theta \sim 1$) limit of (\ref{coinq2}), $\omega(k) \sim \sqrt{k^2+\theta^2}$, that leads to the Dirac equation.  

Despite this quantitative equivalence, there still appears to be two qualitatively distinct approaches to quantizing a random walk.  A possible resolution is simply to consider the CTQW as the discretization of the one-dimensional nonrelativistic Schr{\"o}dinger equation, and the DTQW as the discretization of the one-dimensional Dirac equation.  

The Schr{\"o}dinger equation can be considered the quantization (by analytic continuation) of the diffusion equation for Brownian motion.  The Dirac equation can also be considered the quantization (by analytic continuation) of the two-velocity model for the telegrapher's equation \cite{Gaveau84}.  This model describes a particle that moves with a constant velocity left or right, switching its velocity randomly at some constant rate \cite{Kac74}.  This latter process corresponds precisely to the coined classical random walk originally described above.  From this point of view, the two quantum walks are not two different quantization methods, but rather equivalent quantizations of two different stochastic processes: one described by the diffusion equation and the other by the telegrapher's equation.  That {\it both} lead to propagation with a maximum speed is the surprising yet simple consequence of discretizing equations on a lattice.

I gratefully acknowledge P. R. Johnson and A. J. Dragt for key discussions at the beginning of this work.  

\bibliography{qwalk}

\end{document}